\documentclass[%
reprint,
amsmath,amssymb,
aps,
prl,
]{revtex4-2}
\usepackage{amsmath,bm}
\usepackage{mathtools}
\usepackage{commath}
\usepackage{amssymb}
\usepackage{babel}
\usepackage{cases}
\usepackage[colorlinks=true, citecolor=blue, anchorcolor=green]{hyperref}
\hypersetup{linkcolor=red}
\usepackage{natbib}
\usepackage{floatrow}
\usepackage{dblfloatfix}
\usepackage{xcolor}
\usepackage{lineno}
\usepackage{braket}
\usepackage{graphicx}
\usepackage{dcolumn}
\usepackage{bm}

\begin{document}
	
	\preprint{APS/123-QED}
	
	\title{Photonic Bound States and Scattering Resonances in Waveguide QED}
	
	\author{Bastian Bakkensen}
	\author{Yu-Xiang Zhang}
	\author{Johannes Bjerlin}
	\author{Anders S{\o}ndberg S{\o}rensen}
	\affiliation{%
		Center for Hybrid Quantum Networks (Hy-Q), Niels Bohr Institute, University of Copenhagen, Blegdamsvej 17, 2100 Copenhagen Ø, Denmark
	}%

	\date{\today}
	
	\begin{abstract}
		We study the emergence of two types of two-photon bounds states in waveguides of any chirality.
        Specifically, we present a systematic way of analytically determining the eigenstates of a system consisting of a waveguide coupled to a partially chiral, infinite array of equidistant two-level emitters. Using an effective Hamiltonian approach, we determine the properties  of the two-photon bound states by determining  their  dispersion relation and internal structure. The bound states come in two varieties, depending on the two-photon momentum and emitter spacing.
        One of these states is a long-lived true bound state, whereas the other, a scattering resonance, decays in time via coupling to free two-photon states, leading to resonances and corresponding phase shifts in the photon-photon scattering. 
	\end{abstract}
	
	\maketitle
	
	
	\paragraph*{\label{sec:intro}Introduction.}
Highly-correlated dynamics induced by the electromagnetic force are ubiquitous in  material systems.
In contrast, the inherent interaction between photons, the
carriers of the electromagnetic force, is extremely weak~\cite{Enterria2013}. The interaction can, however, be increased by coupling the photons to material systems and this raises  fundamental questions about how 
to understand effective photon-photon interactions. 
Recently, effective interactions have been achieved in multiple systems including atoms in  cavities~\cite{Hacker:2016wv,turchettemeas}, superconducting circuits~\cite{bishop2009nonlinear}, molecules~\cite{maser2016few},
Rydberg atoms~\cite{Tiarks:2018wi,Thompson:2017wt,jeannic2021,hoffer2016,firstenberg2013attractive,sun2018single}, quantum dots~\cite{Lodahl:2015wq,sun2018single} and nanoscale photonic structures~\cite{Thompson:2013wc,Yu:2019ww}.
This has  enabled experimental observations of exotic photonic states, such as  bound states in a gas of Rydberg atoms~\cite{Liang:2018ty},
strong bunching or antibunching of photon transmitted by atomic ensemble-coupled waveguides~\cite{Prasad:2020tw}  and 
optical transistors gated by a single photon~\cite{Chen:2013wx}.
In addition to their fundamental importance, strong and controllable photon-photon interactions in the 
few-photon regime is also of immense technological importance due to their applications in 
quantum information processing~\cite{Chang:2014ua,duan2001long}.

\begin{figure}
	    \centering
	    \includegraphics[width=0.99\textwidth]{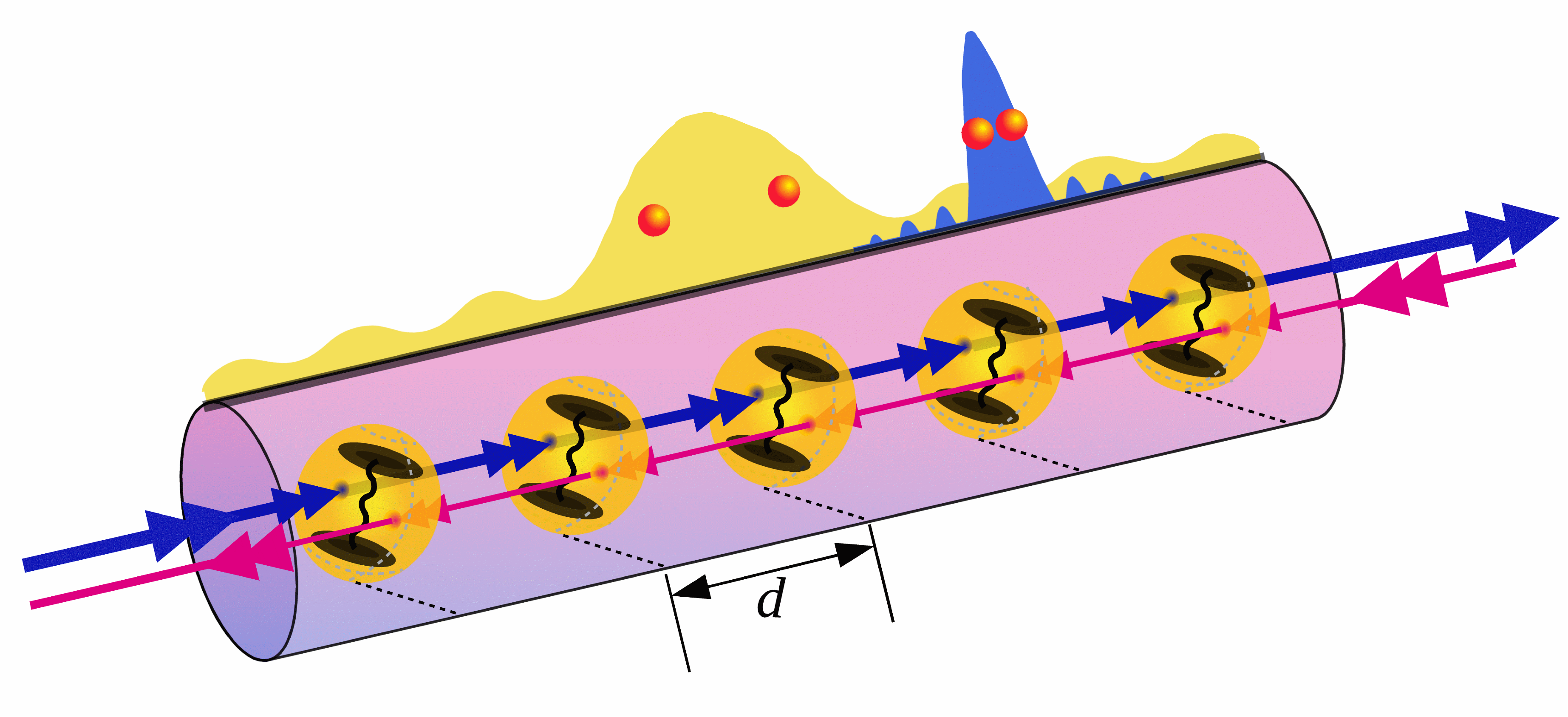}
	    \caption{A waveguide coupled to an infinite 1D array of equidistant two level emitters. We identify two types of solutions for two photons:   true  bound state where the photons bind together indefinitely, illustrated by the blue curve, and scattering resonances,  marked with yellow,  where the photons bind together for a finite time.}
	    \label{fig:system}
	\end{figure}

Perhaps the conceptually simplest avenue for photon-photon interaction is
a waveguide QED setup, where an array of equally spaced two-level emitters are coupled to a one-dimensional (1D)  waveguide~\cite{Roy:2017aa,Sheremet:2021uu}, as illustrated in Fig.~\ref{fig:system}. Despite its simplicity this system exhibits rich and complex physics due to the interplay between scattering by the emitters and effective optical nonlinearities induced by the inability for two photons to excite the same emitter. Detailed understanding of the dynamics has only been obtained for the special case of chiral  interactions,
where the emitters are only coupled to light propagating in one of the two directions~\cite{mahmoodian2020dynamics,yudson1985dynamics,shen2007strongly,shen2007strongly2}.  Several interesting effects including photon bound states have been predicted in this case ~\cite{yudson1985dynamics,shen2007strongly,shen2007strongly2}, but it remains an open question to which degree such phenomena are present in the more generic situation of partially-chiral or
completely non-chiral interactions. 
Understanding these limits is of utmost importance for the experimental exploration of these effects since these only  achieve  partially-chiral or completely non-chiral interactions ~\cite{lodahl2017chiral}.

Recent works have made significant improvements in understanding the 
few-photon sectors of non-chiral waveguide QED, with
predictions of subradiant bound states
~\cite{PhysRevResearch.2.013173,poddubny2020quasiflat}, and other peculiar 
multi-excitation states~\cite{Douglas:2016aa,Zhong:2020uc,Ke:2019aa,Ke2020,Iorsh:2020vd,Lang:2020vo} including free-fermion features
~\cite{Asenjo-Garcia2017,Albrecht2018,Zhang:2019aa,Zhang:2021vd}, quantum Hall phases~\cite{Poshakinskiy:2021wu}, etc.  These approaches have, however, been restricted to finite systems. Although natural for realistic considerations, this also introduces boundary effects that blur the  intrinsic physics of the photon-photon interaction.

In this Letter, we find exact solutions of the two-photon eigenstates of an infinite array of two-level 
emitters coupled to a waveguide with any degree of chirality.
These solutions prove the existence of exact photon bound states  for any degree of chirality.
In addition we identify a new family of eigenstates which are not present in the chiral system. These represent scattering resonances with contributions from both a localised state and free propagating states, as depicted in Fig.~\ref{fig:system}.
Similar \emph{resonance states} have previously only been identified 
for an array of three-level emitters with chiral interactions operated near conditions of 
electromagnetic induced transparency, and were shown to have technological applications for achieving anti-bunched light fields~\cite{iversen2021strongly}. 
Our result shows that such resonances are more ubiquitous  and we provide an intuitive physical picture for the parameter regimes where eigenstates are true bound states or scattering resonances. 

\begin{figure}
    \includegraphics[trim=150 600 150 10,clip,width=0.99\textwidth]{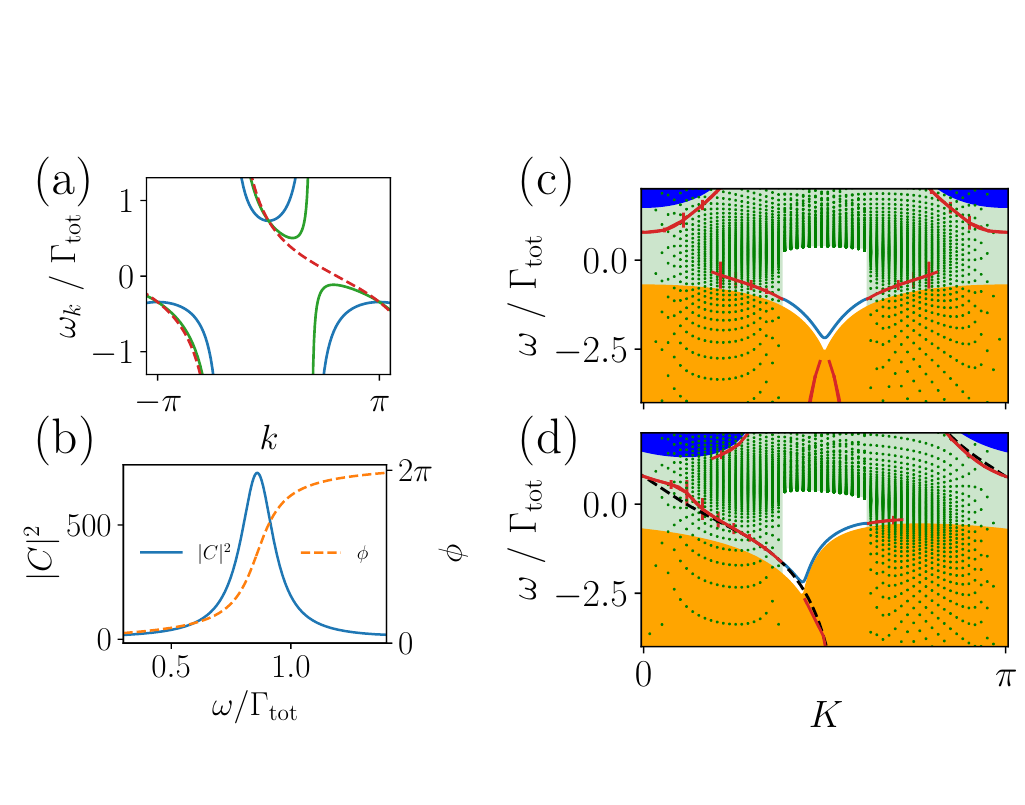}
    \caption{\label{fig:boundset}  (\textbf{a}) Dispersion relation of single photon states for non-chiral $\Gamma_R = \Gamma_L$ (blue), partially chiral $\Gamma_R = 0.1 \Gamma_{\textrm{tot}}$ (green) and chiral $\Gamma_R = 0$ (red). 
    As the system approaches the chiral regime, the band gap narrows until it completely disappears for chiral systems. (\textbf{b}) Amplitude and phase of the scattering resonance for $\Gamma_L=\Gamma_R$,  and $K = 0.2$. The phase switches at the peak of the resonance.   (\textbf{c}) Dispersion relation of two-photon bound states for non-chiral systems $\Gamma_R = \Gamma_L$.  The shaded blue (orange) area indicates the energy range with a continuum of two free states from the upper (lower) branch of the single excitation dispersion relation. The shaded green area indicates the region accessible by photon states combining upper and lower branch and green dots indicate the density of state in a given area. Lines in the unshaded region are the dispersion relation or the bound states. Lines in the shaded area indicate the resonance states with widths indicated by the bars. (\textbf{d}) Same as (c), but for $\Gamma_R = 0.25 \Gamma_{\textrm{tot}}$. The dashed black line is the dispersion relation for the bound state in the completely chiral case, $\Gamma_R = 0$. As the system becomes more chiral, the branches merge together on one side of the gap, approaching the chiral dispersion relation. 
    All graphs are for $k_0 = 1.2$.}
    \end{figure}

	The main results of this Letter are shown in Fig. \ref{fig:boundset}. For single photons, the dispersion relation   exhibits a gap, except for the completely chiral case, as shown in Fig.~\ref{fig:boundset}(a). This leads to a similar gap for certain momenta in the spectrum of two free single photons: In Fig.~\ref{fig:boundset}(c,d) the shaded regions show the possible energies obtained by combining the single particle energies, $\Omega(K, q) = \omega_{K + q/2} + \omega_{K - q/2}$, where $K$ is the center of mass momentum per photon, $q$ is the relative momentum between the two photons and $\omega_k$ is the energy of a single photon state with wave number $k$.
	In the gapped region we find two-photon bound states. Outside this region the states become scattering resonances as illustrated in Fig.~\ref{fig:boundset}(b).  These resonances come in a number of different branches. For the nonchiral case  shown in Fig.~\ref{fig:boundset}(c), one of these branches connects to the previously identified bound state at $K=0$ \cite{PhysRevResearch.2.013173}, where the lifetime becomes infinite because the density of states vanishes. A second branch is located in the vicinity of the gapped region and connects to the bound states. The third branch appears in the continuum beneath the gapped region and corresponds to the continuation of left and right moving bound states from the gapped region.
	As the system becomes more chiral, these branches start to merge in nontrivial ways, beginning to align with the chiral bound state dispersion, as illustrated in Fig.~\ref{fig:boundset}(d).
	
	\paragraph*{\label{sec:model}Model.}
	We consider an infinite array of two-level emitters separated by a distance $d$.
    Taking the Born-Markov approximation, we eliminate the waveguide modes while retaining their effects in an effective Hamiltonian, $H_{\textrm{eff}}$, for the emitters \cite{caneva2015quantum},
	\begin{align}\label{heff}
	\begin{split}
	H_{\textrm{eff}} = 
	&-i \Gamma_R \sum_{j<i} e ^{ik_0 |i - j|d} \sigma_i^{\dagger} \sigma_j -i \Gamma_L \sum_{i<j} e ^{ik_0 |i - j|d} \sigma_i^{\dagger} \sigma_j \\
	& -i \sum_{i=-\infty}^{\infty} \frac{\Gamma_R + \Gamma_L}{2}  \sigma_i ^{\dagger} \sigma_i,
	\end{split}
	\end{align}
	where $\Gamma_L$ ($\Gamma_R$) is the rate of decay from an individual emitter to the left (right) propagating modes, $k_0$ is the wave number resonant to the emitter transition, and $\sigma_i^{\dagger}$ ($\sigma_i$) is the spin creation (annihilation) operator for an excitation of the $i$'th emitter. The distance $d$ is set to $d=1$ throughout the remaining text with the understanding that all wavenumbers $k$ and $k_0$ denote $kd$ and $k_0d$. The coupling is non-chiral if $\Gamma_R = \Gamma_L$, partially chiral if $\Gamma_R\neq\Gamma_L$ and chiral if one of the coupling rates is zero. The chirality can be quantified by $\Gamma_R/\Gamma_{\textrm{tot}}$ with $\Gamma_{\textrm{tot}}=\Gamma_R+\Gamma_L$. 
	
	For the single-photon eigenstates,  translation invariance implies that
	they have the form of Bloch states
	$\ket{k}=\sum_j e^{ikj}\sigma_j^\dagger \ket{0}$ with $k\in[-\pi, \pi]$, i.e., within the Brillouin zone.
	The eigenenergy of $\ket{k}$ is found to be 
	\begin{equation}
	\omega_k = \frac{\Gamma_R}{2}\cot\frac{k_0 - k}{2} + 
	\frac{\Gamma_L}{2} \cot\frac{k_0 + k}{2}.
	\label{eq.singleenergy}
	\end{equation}
	This is plotted for three different levels of chirality in Fig.~\ref{fig:boundset}$(a)$. A notable feature is the energy gap between the connected bands separated by the singularities at $k=\pm k_0$ (the superradiant modes)~\cite{Zhang:2019aa}, unless the system is completely chiral. This feature is crucial for the two-photon bound states.

	For two-photon states, we exploit the 
	translation invariance by working with  basis states
	\begin{equation}
	| K, \Delta \rangle = \sum_x e^{2iKx} \sigma_{x - \Delta/2}^{\dagger} \sigma_{x + \Delta/2}^{\dagger} | 0 \rangle 
	\end{equation}
	which are  eigenstates of both center of mass momentum ($2K$) and relative distance $\Delta$. 
	Again, translation invariance prevents the couplings between states of different momentum $K$. We
	therefore restrict $H_{\textrm{eff}}$ to the subspace of fixed $K$ and write it
	as $H_{\textrm{eff}}^K = \sum_{\Delta, \Delta'} \mathcal{H}_{\Delta \Delta'}^K |K, \Delta \rangle \langle K, \Delta'|$ where
	\begin{align}
	\begin{split}
	\mathcal{H}_{\Delta, \Delta'}^K = &-i\Gamma_R \sum_{\varepsilon = \pm 1} e^{i(k_0 - K) |\Delta + \varepsilon \Delta'|}\\ &-i\Gamma_L \sum_{\varepsilon = \pm 1} e^{i(k_0 + K) |\Delta + \varepsilon \Delta'|}.
	\label{eq.hamelements}
	\end{split}
	\end{align}
	This Hamiltonian reduces the ``two-body'' problem to a ``one body'' one. This paves the way for a clean analytical solution to the bound and resonance states. 
	
	\begin{figure}
		\centering
		\includegraphics[trim=142 600 115 65,clip,width=0.99\textwidth]{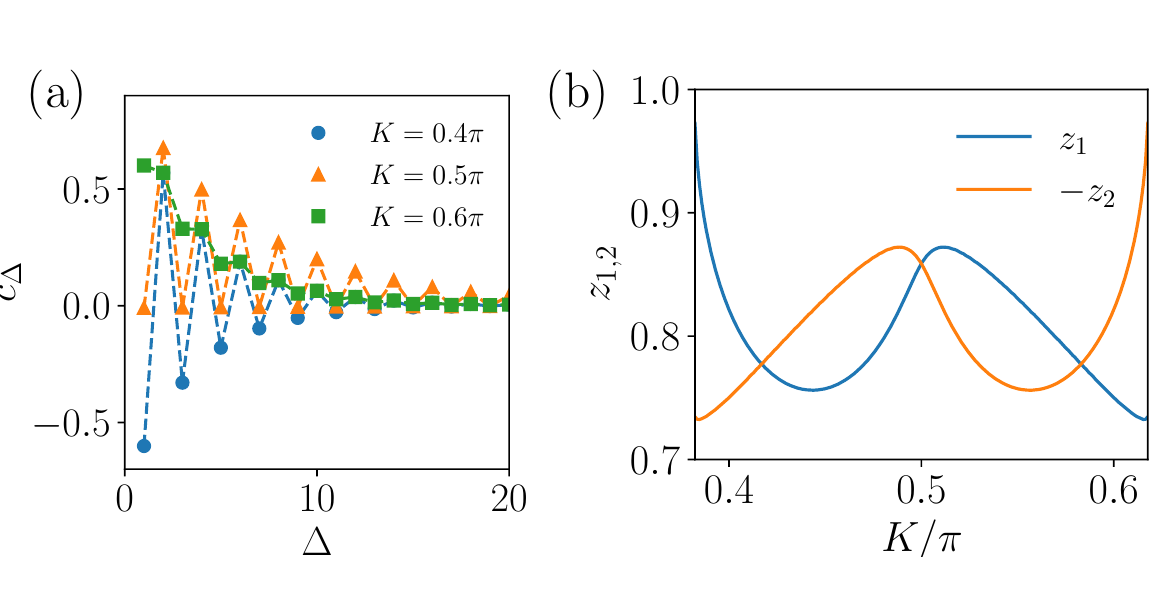}
		\caption{\label{fig:eigrel} (\textbf{a}) Three examples of 
		two-photon bound states in the gapped region for non-chiral interactions with $k_0 = 1.2$. (\textbf{b}) Value of $z_{1,2}$ in the gapped region for the non-chiral case. As one approaches the edges, one of the two values tends towards  1, indicating the transition to a scattering resonance  at the boundary. Here $k_0 = 1.2$.}
	\end{figure}

	\paragraph*{\label{sec:ansatz}Analytical Solutions.}
	
	To obtain the eigenstates of $H^{K}_{\textrm{eff}}$, we consider the following ansatz
	\begin{equation}
	| z \rangle = \sum_{\Delta = 1}^{\infty} z^{\Delta} | \Delta \rangle.
	\end{equation}
	Here $z$ is a complex number that must satisfy $\abs{z}\leq 1$ to have physical meaning. We here use the abbreviation $\ket{\Delta} \equiv \ket{K, \Delta}$. 
	For $\abs{z}< 1$, $\ket{z}$ decays exponentially with $\Delta$, hence is expected to capture the properties of bound states. $\ket{z}$ with $\abs{z}=1$ is a spin-wave like Bloch state representing free particles. 
	
	Applying $H^{K}_{\textrm{eff}}$ to $\ket{z}$ yields 
	\begin{equation}
	H^{K}_{\textrm{eff}} |z \rangle = \omega^{K}(z)|z \rangle + ig_+ (z) | z^{0}_{+} \rangle + ig_-(z) | z^{0}_{-} \rangle
	\label{eq.almosteigen}
	\end{equation}
	where $z^{0}_{\pm} = e^{i(k_0 \pm K)}$ and
	\begin{align}
	\begin{split}
	\omega^K(z) = & \frac{2z \Gamma_L \sin(k_0 + K)}{1 + z^2 - 2z\cos(k_0 + K)}  \\ + & \frac{2z \Gamma_R \sin(k_0 - K)}{1 + z^2 - 2z\cos(k_0 - K)}.
	\end{split}
	\end{align}
	Expressions for $g_{\pm}(z)$ are 
	given in the Supplemental Material~\cite{sp}, but we note that both functions are real. 
	
	It is important to realize that $\ket{z^{0}_{\pm}}$ is independent of 
	$z$. By exploiting this feature, we can find eigenstates of
$H^{K}_{\textrm{eff}}$ with eigenvalue $\omega$  by  first collecting roots $\{z_i\}$ satisfying $\omega^K(z_i)=\omega$, and then finding a proper set of coefficients $\{a_i\}$ so that the superposition
$| \psi \rangle = \sum_{i} a_i | z_i \rangle$ is the desired eigenstate.
Applying the Hamiltonian to $\ket{\psi}$ yields
	\begin{equation}
	H^{K}_{\textrm{eff}} |\psi \rangle = \omega | \psi \rangle + i\sum_{\epsilon=\pm}\sum_{i} a_i  g_\epsilon(z_i)|z^{0}_{\epsilon} \rangle.
	\label{eq:cancel}
	\end{equation}
	Thus $\ket{\psi}$ is the desired  eigenstate if $\{a_i\}$ fulfills \begin{equation}
	    \sum_i a_i g_{\epsilon}(z_i)=0,
	    \label{eq:plusandminus}
	\end{equation}
	for $\epsilon=\pm$ simultaneously.
	
	To proceed, we note that the size of the root set $\{z_i\}$ is exactly four  because $\omega^K(z)=\omega$ can be reduced to a polynomial equation of degree four, see
	the Supplemental Material~\cite{sp}. Moreover, it can be verified that if $z$ is a solution, so is $1/z$. Since roots with $\abs{z}>1$ must be excluded from the ansatz 
	the two-photon states can be grouped  according to the number
	of roots with magnitude less than one. This leads to three families of state: (1) the scattering states, (2) the bound states, and (3) the resonance states.
	
	\paragraph*{The scattering states.}  If all of the four roots of
	$\omega^K(z)=\omega$ have unit magnitude $\abs{z}=1$ for a given $K$ and $\omega$, the superposition of them, $\ket{\psi}=\sum_i a_i \ket{z_i}$, is  a scattering eigenstate \cite{schrinski2021}. Note that 
	Eq.~\eqref{eq:plusandminus} only provides two constraints but we have three independent parameters (recall the normalization of states). Hence nontrivial
	solutions for the coefficients $\{a_i\}_{i=1}^{4}$ typically exist. Such scattering states form the majority of eigenstates, but are not the focus of this Letter.
	
	\paragraph*{The bound states.} If the equation $\omega^K(z)=\omega$ has no roots of unit magnitude, but two roots $z_{1,2}$ satisfying $\abs{z_{1,2}}<1$ (the conjugate roots $z_{1,2}^{-1}$ are excluded since they have magnitudes larger than one), the corresponding eigenstate consists entirely of localized states and  is thus a bound state
	\begin{equation}
	|\psi \rangle = A|z_1\rangle + B |z_2\rangle.
	\end{equation}
According to Eq.~\eqref{eq:plusandminus} solutions for $A$ and $B$ exist if and only if $z_{1,2}$ fulfill
	\begin{equation}
	g_- (z_1) g_+(z_2) - g_+(z_1) g_-(z_2) = 0
	\end{equation}
We can now determine the dispersion relation of the bound states by searching for frequencies $\omega$ fulfilling this relation for a certain value of $K$. 
We find such bound states in the regime $k_0 < K < \pi - k_0$ for $k_0 < \pi / 2$ and $k_0 > K > \pi - k_0$ for $k_0 > \pi / 2$. The solution in the former regime is show in 
 Fig~\ref{fig:boundset}(c,d).
As shown in the figure, the bound states appear in an energy gap within the
spectrum of free particle states. 
This provides a simple explanation for the stability of the bound states: there are simply no scattering states that match in both energy and momentum.


	For the bound state $z_{1,2}$ are always real but have opposite signs. Thus, the bound
	states have real wavefunctions when expressed in terms of $\Delta$.  We plot the wavefunctions for a few  bound states in Fig. \ref{fig:eigrel}(a). These clearly shows that the amplitude decays exponentially with
	increasing $\Delta$.
	As the bound state approach the border of the gapped region, 
we find that $|z_{1,2}| \rightarrow 1$ for one of the two values, as shown in Fig. \ref{fig:eigrel}(b). This means that one of the components becomes delocalized and signifies the cross-over into a delocalized resonance state.  
	
In addition to the bound states in the gap, we find two different types of bound states.
	 For $K=0$ or $K=\pi$ we find a bound state given by $\ket{z=\cos(k_0)}$.
	As opposed to those protected  by energy gaps, this bound state has a different physical interpretation: it appears because at this particular $K$ the density of  states for  two free photons vanishes.
	Another exception is found in the chiral limit. Here the bound states are given by  $\ket{z}$ with $z=\cos(k_0 \pm K)$, corresponding to the bound states identified in Ref.~\cite{yudson1985dynamics,shen2007strongly,shen2007strongly2}. Here the plus sign is for $\Gamma_R = 0$ and minus for $\Gamma_L = 0$. 
	
	\begin{figure}
		\centering
		\includegraphics[width=0.99\textwidth]{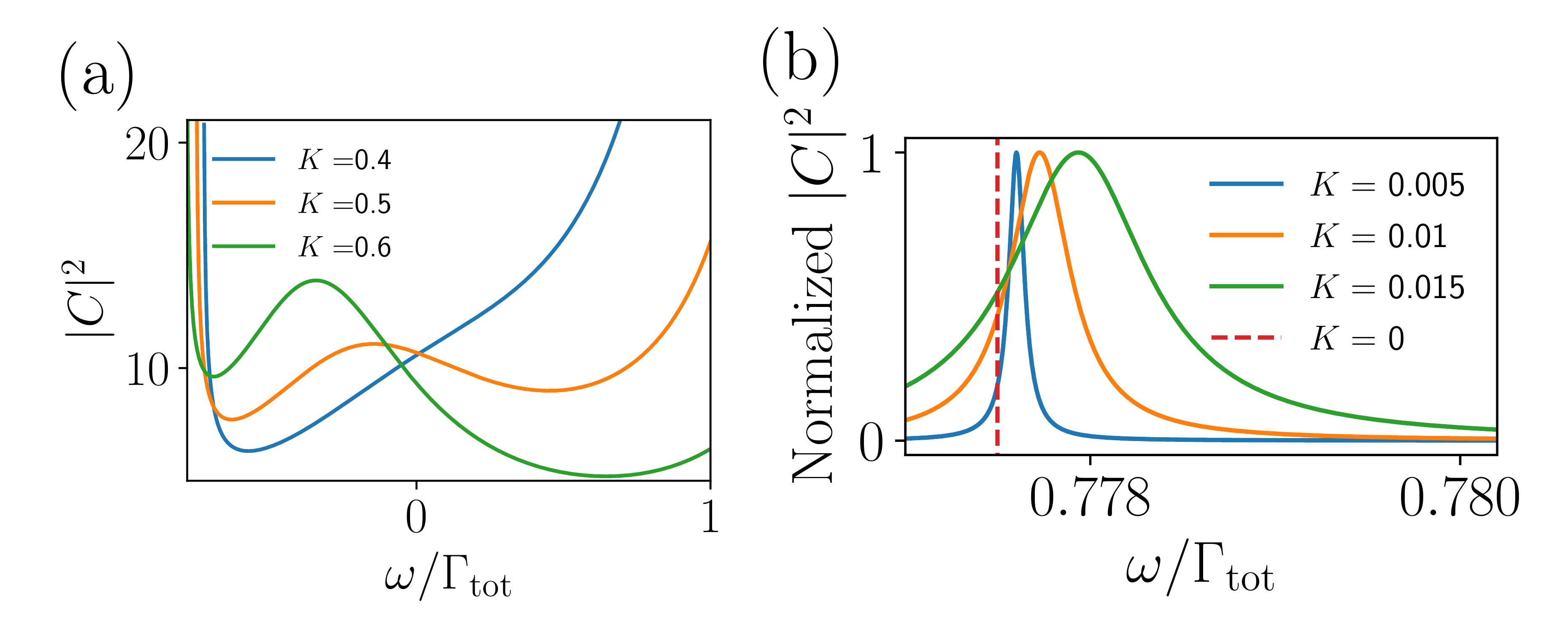}
		\caption{(\textbf{a}) Disapperance of the branch connecting to the gapped region. The peak  is a well-defined for $K$ near the gapped region, but disappear as $K$ decreases, rendering the resonance ambiguous.
		(\textbf{b}) Resonances in the vicinity of $K = 0$.  As $K\rightarrow 0$ the peak of the resonance converges the theoretical value at $K = 0$, marked with a dashed line. The width of the resonance increases linearly with $K$. All curves have been normalized to a peak value of 1.
		For both plots $\Gamma_L / \Gamma_R = 1$ and $k_0 = 1.2$.}
		\label{fig:resonanse}
	\end{figure}

\paragraph*{The resonance states.}
A different  family of states is obtained if  
$\omega^K(z)=\omega$ has a single solution, $z_b$, with $\abs{z_b}<1$ (the conjugate root $z_b^{-1}$ 
is excluded) and two roots of unit magnitude denoted by $e^{\pm iq}$. 
In this case, it is impossible to construct an eigenstate out of just two of the three states.
By combining  all  three states 
we are, however, always able
to find a proper superposition that fulfills the two restrictions imposed by Eq.~\eqref{eq:plusandminus}. 
 These solutions we write in the form 
 \begin{equation}
 | \psi \rangle = C|z_b\rangle_{\textrm{N}} + |e^{iq}\rangle + e^{i\phi} | e^{-iq} \rangle
 \end{equation}
where $|z_b\rangle_{\textrm{N}} = \sqrt{|z_b|^{-2}-1} |z_b\rangle$ is the normalized version of $|z_b \rangle$. $C$ is a factor that measures how much weight the localized state has relative to  the propagating states and $\phi$ is a phase shift. 
When we vary $\omega$ we see a clear peak in   $|C|^2$ and a change of in $\phi$ by $2\pi$ around the peak, as shown in  Fig ~\ref{fig:boundset}(b). This behaviour matches a scattering resonance, and we shall therefore refer to these states as resonance states. 

For any given $K$, the range in energy bounded by
$\omega^K(\pm 1)$, where $\abs{z_b}<1$, allows resonance states.
In Fig.~\ref{fig:boundset}(c,d) we show the location of the peaks of $|C|^2$ and also mark the width of the peaks.  
These resonances  come in a number of different branches. For concreteness we discuss these for the non-chiral case in Fig.~\ref{fig:boundset}(c). 

 %
 
 One branch of  resonances connects to the bound state at $K=0$. For these states the width of the resonance approach zero for $K\rightarrow 0$, see Fig. \ref{fig:resonanse} (b). The energy of this branch diverges
for  $K\rightarrow k_0$ (and $\pi-k_0$), 
 approaching the lower border of the blue region in Fig.~\ref{fig:boundset}(c). In this limit, we find that $z_b$ tends towards unity while the width goes to zero. These scattering resonances thus become weakly localized but long-lived. 

	Another  branch directly connects to the bound states in  the gapped region at $k_0$.  This branch displays clear features of scattering resonances, with a  width going to zero, when $K$ is close to the bound state regime. However, this feature becomes ambiguous when 
    $K$ moves further away. To illustrate this, we focus on the 
    region $K<k_0$. 
    We  consider three examples  $K=0.4, 0.5,$ and $0.6$ with $k_0 = 1.2$ and  plot
$|C|^2$ as a function of $\omega$ in Fig. \ref{fig:resonanse} (a).
For $K = 0.6$, $|C|^2$ exhibits a peak. 
As $K$ decreases, however, the peak becomes increasingly ill-defined, until it has completely disappeared at $K = 0.4$, illustrating the fading of the resonance states in Fig. \ref{fig:boundset}(c).
Additionally, Fig.~\ref{fig:resonanse} (a) also shows a divergence of $\abs{C}^2$ near the edge. This divergence is accompanied by 
$\abs{z_b}\rightarrow 1$, and  hence does not signify strong binding.

A third type of branches appears in the region below the gap. If we consider the  stable bound states to have a left and a right moving branch, these resonances represent the continuation  of each of these branches. Unlike the branch discussed above, these do not connect to the  bound state, but fade away as they approach.
	

	At last, we consider the situation as we approach 
	the chiral limit. The chiral case is special also for the single-photon states. As shown in Fig. \ref{fig:boundset}(a) the upper and lower bands of the single particle states tend to merge together when the chirality increases and finally close the gap in the chiral limit.
	Consequently, the different two-photon branches also tend to merge and the
	 resonances become narrower, until finally they reduce to
	the bound states of the chiral limit, see Fig.~\ref{fig:boundset}  (d).
Seen from this perspective, the non-chiral limit [Fig.~\ref{fig:boundset}  (c)] can also be understood as the interplay between a left and a right moving branch from the chiral case.  
	
	\paragraph*{\label{concl} Conclusion and Outlook.}
	
	We have identified the essential physics  of photon-bound states for emitters of arbitrary chirality. The bound state solutions come in two categories: true photonic bound states and scattering resonances. The latter represent a novel phenomena for two-level systems, not present in the chiral case. Our extension to the non-chiral regime paves the way for  explorations of photon bound states  in a much  wider  range of physical system, including, e.g. the microwave regime with superconducting systems. Furthermore, our results  represents the first steps towards more general explorations of  many-body dynamics of photons, e.g. strongly interacting photon fluids.
	
		\paragraph*{Note added.}
	During the final stages a related preprint reported numerical results reminiscent of the result reported her~\cite{GDnumerics}.

	\begin{acknowledgments}
	We are grateful to Bj{\"o}rn Schrinski, Sahand Mahmoodian, Thomas Pohl,   Giuseppe Calajo, Darrick E. Chang and Klemens Hammerer for useful discussion.  We acknowledge the support of Danmarks Grund-
forskningsfond (DNRF 139, Hy-Q Center for Hybrid
Quantum Networks). 
	\end{acknowledgments}
	
	
	\bibliography{two_photon_bound}
	
	\pagebreak
	\widetext
	\begin{center}
		\textbf{\large Supplemental Materials: Photonic Bound States and Scattering Resonances in Waveguide QED}
	\end{center}
	\setcounter{equation}{0}
	\setcounter{figure}{0}
	\setcounter{table}{0}
	\makeatletter
	\renewcommand{\theequation}{S\arabic{equation}}
	\renewcommand{\thefigure}{S\arabic{figure}}
	\renewcommand{\bibnumfmt}[1]{[S#1]}
	\renewcommand{\citenumfont}[1]{S#1}

	\section*{Solving $\omega^K$ as a Quartic Equation}
	\label{sec.quartic}
	
	The necessary functions defined in the main text have the explicit forms
	
	\begin{align}
	\begin{split}
	\omega^K(z) = & \frac{2z \Gamma_L \sin(k_0 + K)}{1 + z^2 - 2z\cos(k_0 + K)} + \frac{2z \Gamma_R \sin(k_0 - K)}{1 + z^2 - 2z\cos(k_0 - K)} \\ g_+(z) = & 2 z \Gamma_L  \frac{z - \cos(k_0 + K)}{1 + z^2 - 2z\cos(k_0 + K)} \\ g_-(z) = & 2 z \Gamma_R  \frac{z - \cos(k_0 - K)}{1 + z^2 - 2z\cos(k_0 - K)}
	\label{eq:omg}
	\end{split}
	\end{align}
	
 We are able to recast the equation $\omega^K(z)=\omega$ as a quartic polynomial equation 
 by multiplying both sides by the two denominators. It leads to
 a quartic equation in the general form of
	\begin{equation}
	c_4z^4 + c_3z^3 + c_2z^2 + c_1z + c_0 = 0
	\end{equation}
	where the coefficients are 
	\begin{align}
	\begin{split}
	c_4 & = c_0= \omega, \\ 
	c_3 & = c_1= -2\bigg[\omega(c_+ + c_-) + \Gamma_L s_+ + \Gamma_R s_-\bigg], \\ 
	c_2 & = 2\bigg[\omega(1 + 2c_+ c_-) + 2 (\Gamma_L s_+c_- + \Gamma_R s_- c_+)\bigg], 
	\end{split}
	\end{align}
	with $s_{\pm} = \sin (k_0 \pm K)$ and $c_{\pm} = \cos(k_0 \pm K)$. 
	The fundamental theorem of algebra implies that a quartic equation has four solutions. 
	Formulae for these solutions are well known for almost four centuries.  
	
	The symmetry of the coefficients implies that if $z_1$ is a solution, 
	so is $z^{-1}_1$. Therefore, the four solutions can be denoted by $z_{1,2}$ and $z^{-1}_{1,2}$. We define $z_1+z_1^{-1}=u_1$ and $z_2+z_2^{-1}=u_2$, and obtain that
	\begin{equation}
	    \begin{aligned}
	        u_1+u_2 & =-c_3/\omega \\
	        u_1 u_2 & =c_2/\omega-2
	    \end{aligned}
	\end{equation}
	Then it is straightforward to firstly solve for $u_{1/2}$, and hence for $z_{1/2}$ as well.

\end{document}